\documentclass[]{raa}
\usepackage{graphicx,times}
\usepackage{amsmath}
\usepackage{natbib}
\usepackage{amssymb,amsmath}
\bibpunct{(}{)}{;}{a}{}{,}

\usepackage[a4paper=true,dvipdfm=true,pagebackref=true]{hyperref}
\hypersetup{pdftitle = The title of my PDF, pdfauthor = My name, pdfsubject= The subject, pdfkeywords = keyword1 keyword2 keyword3}
\hypersetup{colorlinks = true, linkcolor = green, anchorcolor = red, citecolor = blue, filecolor = red, pagecolor = red, urlcolor = red}

\begin{document}

 \title{3-D kinematics of classical Cepheids according to Gaia\,EDR3 catalog}
 \volnopage{ {\bf 20XX} Vol.\ {\bf X} No. {\bf XX}, 000--000}
   \setcounter{page}{1}
   \author{V. V. Bobylev, A. T. Bajkova}
   \institute{Pulkovo Astronomical Observatory, St.-Petersburg 196140, Russia;
   {\it vbobylev@gaoran.ru}\\
   \vs \no
   {\small Received 20XX Month Day; accepted 20XX Month Day}
 }

\abstract{
The kinematics of about 2000 classical Cepheids of the Milky Way with data from Gaia\,EDR3 catalog has been studied. For some of these stars, there are line-of-sight velocities. On the basis of the nonlinear rotation model, the parameters of the rotation curve of the Galaxy were determined. The circular linear rotation velocity of the near-solar neighborhood around the Galaxy center was $V_0=236\pm 3$~km s$^{-1}$ for the assumed Sun's galactocentric distance $R_0=8.1\pm0.1$~kpc. Analysis of residual velocities of Cepheids based on the linear Ogorodnikov-Milne model showed the presence of the following significantly different from zero gradients: $\partial U/\partial x,$ $\partial U/\partial z,$
$\partial V/\partial x,$ $\partial V/\partial z$ and $\partial W/\partial x,$ which behaves differently depending on the selection radius. The most interesting is the gradient $\partial W/\partial x\sim-0.5\pm0.1$~km s$^{-1}$ kpc$^{-1}$ (positive rotation of this star system around the galactic axis $y$, $\Omega_y$) since the velocities $W$ are free of galactic rotation. Here we have an indirect influence of various effects leading to a perturbation of the vertical velocities of the galactic disk stars. Based on a simpler model, a more accurate estimate of this rotation is obtained, $\Omega_y=0.51\pm0.07$~km s$^{-1}$ kpc$^{-1}$.
}

   \authorrunning{V. V. Bobylev, A. T. Bajkova}    
   \titlerunning{3-D kinematics of classical Cepheids}  
   \maketitle

 \section{INTRODUCTION}
Studying the rotation of the Galaxy is an important astronomical
task. For this purpose, various galactic objects are used~---
stars, star clusters, hydrogen clouds, star-forming regions, maser
sources, etc. The local parameters of galactic rotation are best
studied in the region, where there are massive measurements of
high-precision distances and kinematic characteristics of stars.
At present, the radius of such a circumsolar region is 2--4~kpc.
There are also other precise studies that measure rotation curves
for heliocentric distance larger than 4~kpc \citep{Xue2008,Eilers2019,Reid2019,Ablimit20,Bob2021,Drimmel2022}.

Cepheids make it possible to realize an independent distance scale due to the use of the period--luminosity relation \citep{Leavitt1908,Leavitt1912}. Modern calibrations of this ratio allow estimating distances to Cepheids with errors less than 10\% \citep{Berd00,Sandage06,Skowron19a}. Using such distances in combination with high-precision proper motions of about 220 classical Cepheids from the catalog~\cite{Hip1997}, there were refined, for example, the Galaxy rotation parameters~\citep{Feast97, Melnik2015}, the spiral structure parameters~\citep{Melnik1999, BobylevBaj12, Dambis15} and the distance to the center of the Galaxy (\cite{Chen18}). According to data on older type II Cepheids, there were refined the parameters of the central bulge and the distance to the center of the Galaxy as well (\cite{Majaess09, Bhardwaj17}.

Recently~\cite{Skowron19a}, published an extensive catalog with new data on classical Cepheids. It contains more than 2\,200 stars, which are located in an area around the center of the Galaxy with a radius of about 24 kpc. According to these authors, the distances to these Cepheids are determined using the period--luminosity ratio with errors of 5--10\%. Thus, these Cepheids provide a unique opportunity to study the kinematic properties of the Galaxy in a very wide range of distances. Using the distances to these Cepheids, with the addition of their proper motions from the Gaia~DR2 catalog~\citep{Prusti16,Brown18}, a number of works devoted to the analysis of the spatial distribution of Cepheids~\citep{Skowron19a,Skowron19b,BB2021a}, the Galaxy kinematics~\citep{Mroz19,Ablimit20,BB2021b}, and its spiral structure (for example, \cite{Bob2021}) have been performed.

In the Milky Way galaxy, there is a large-scale wave-like deformation of a thin disk~-- a galactic warp. The positions of the stars have characteristic deviations from the galactic plane, which was established from an analysis of the distribution of various objects: neutral~\citep{Westerhout57,Kalberla2008} and ionized hydrogen \citep{Russeil03,Cersosimo09}, pulsars \citep{Yusifov04}, OB stars \citep{Miyamoto1998,Drimmel00}, red clump and old stars~\citep{Lopez-Corredoira2002,Momany06,Wang2020}, interstellar dust~\citep{Drimmel2001}, or by Cepheids~\citep{Fernie1968,Berd-1987,Bob2013a,Skowron19b}.

Attempts to assess the kinematic properties of the warp and its influence on the motion of stars were undertaken in the works of a number of authors (for example, \cite{Miyamoto1998,Drimmel00,Bob2013b,Poggio2018,Poggio2020,Wang2020,BB2021b}). In particular, \cite{Poggio2020}, from an analysis of about 12 million giants from the Gaia\,DR2 catalog, estimated the warp precession rate as $10.86\pm0.03$ (stat.)~$\pm3.20$  (syst.) km s$^{-1}$ kpc$^{-1}$ in the direction of rotation of the Galaxy.
According to~\cite{Chrobakova21}, the value found by~\cite{Poggio2020} greatly exceeds theoretical predictions. \cite{Chrobakova21} showed that there is no need for precession if the change in the deformation amplitude depending on the age of the analyzed stellar population is taken into account. With this allowance, they, in particular, obtained an estimate of the deformation precession $4^{+6}_{-4}$~km s$^{-1}$ kpc$^{-1}$.
From the line-of-sight velocities and proper motions of classical Cepheids from the Gaia\,DR2 catalog, \cite{BB2021b} found residual rotation (as a manifestation of the warp effect) around the galactic axis $y$ with a velocity $\Omega_y=0.54\pm0.15$~km s$^{-1}$ kpc$^{-1}$ and positive rotation around the $x$ axis with a velocity $\Omega_x=0.33\pm0.10$~km s$^{-1}$ kpc$^{-1}$.

Currently, there are a number of papers devoted to discussing not only Warp, but also other effects associated with vertical oscillations in the disk of the Galaxy \citep{Antoja2018,Kumar2022,Bland-Hawthorn2021,Yang2022}.

The accuracy of parallaxes and proper motions of stars is important for solving a wide variety of kinematic problems. Currently, a version of the Gaia\,EDR3 catalog (Gaia Early Data Release~3, \cite{Brown21}) has appeared, in which, in comparison with the previous version, Gaia DR2~\citep{Brown18}, the trigonometric parallax values and proper motions are refined by about 30\%  for about 1.5 billion stars.

In the work of \cite{BB2021b}, using the proper motions of classical Cepheids from the Gaia\,DR2 catalog based on the linear Ogorodnikov-Milne model, it was shown that there are the following gradients significantly different from zero: $\partial U/\partial z,$ $\partial V/\partial z$ and $\partial W/\partial x,$ which behaves differently depend on the selection radius. These gradients are closely related to the influence of the warp. In this work, due to the greater accuracy of the proper motions used, we obtain more accurate estimates of the values of these gradients, more accurate estimates of the residual rotation rates $\Omega_y$ and $\Omega_x$.

The aim of this work is to study the three-dimensional kinematics of the classical Cepheids of the Milky Way. For this, we supplied the Cepheids from~\cite{Skowron19a} with proper motions from the Gaia\,EDR3 Catalog. We apply a variety of kinematic models for the rotation of the Galaxy. Our task is to estimate the parameters of galactic rotation, as well as to identify the kinematic effects associated with the warp. The main interest of this work is related precisely to the study of the influence of the warp on the kinematics of the Cepheids. Therefore, we are doing everything to use as many of the most distant Cepheids as possible.

The work is structured as follows. An overview of the problem is given in the introduction~\ref{intro}. Section~\ref{methods} describes the nonlinear method for determining the parameters of the Galactic rotation curve~\ref{rotat}, as well as the linear Ogorodnikov-Milne model~\ref{ogor-milne}. Data on Cepheids is described in Section~\ref{data}. The method for generating residual velocities of Cepheids is given in Section~\ref{residual}. The discussion of the values of the found local parameters ($(U,V,W)_\odot$ and $|V_0|$) in Sections~\ref{local} and \ref{rotat-parameters}, the parameters of the galactic rotation curve is given in Section~\ref{rotat-parameters}, and warp-related parameters in Sections~\ref{warpYZ} and \ref{warpXZ}.

 \section{METHODS} \label{methods}
From observations we have the following values: right ascension and declination~--- $\alpha$ and $\delta$, proper motions in right ascension and declination~--- $\mu_\alpha\cos\delta$ and $\mu_{\delta}$, line-of-sight velocity $V_r$. From $\alpha$ and $\delta$ we go to galactic longitude and latitude $l$ and $b$; the heliocentric distance $d$ for Cepheids is calculated on the basis of the period--luminosity relation; we translate the observed proper motions into proper motions in the galactic coordinate system~--- $\mu_l\cos b$ and $\mu_b$. As a result, we have 3 components of the spatial velocity of the star: $V_r$ and two projections of the tangential velocity~--- $V_l=k\,d\,\mu_l\cos b$ and $V_b=k\,d\,\mu_b$, where $k=4.74$~km s$^{-1}$, and $V_r,\,V_l,\,V_b$ are expressed in km s$^{-1}$ (proper motions are given in mas/year (milliarcseconds per year), and the heliocentric distance~--- in kpc).

In this work, we use a rectangular galactic coordinate system with axes directed from the observer towards the galactic center ($x$ axis or axis~1), in the direction of galactic rotation ($y$ axis or axis~2) and in the direction of the North Pole of the Galaxy ($z$ axis or axis~3).

\subsection{Galaxy Rotation Curve} \label{rotat}
To determine the parameters of the galactic rotation curve, we use the equations obtained from the Bottlinger formulas, in which the angular velocity $\Omega$ is expanded in a series up to terms of the $i$-th order of smallness $d/R_0:$
\begin{equation}
 \begin{array}{lll}
  V_r=-U_\odot\cos b\cos l-V_\odot\cos b\sin l-W_\odot\sin b\\
      -R_0\sin l\cos b\left[\sum\limits_{i=1}^N(R-R_0)^i
      {\displaystyle\Omega_0^{(i)}\over \displaystyle i!}\right],
 \label{EQ-11}
 \end{array}
 \end{equation}
 \begin{equation}
 \begin{array}{lll}
   V_l=U_\odot\sin l-V_\odot\cos l-d\Omega_0\cos b\\
 -(R_0\cos l-d\cos b)\left[\sum\limits_{i=1}^N(R-R_0)^i
 {\displaystyle\Omega_0^{(i)}\over \displaystyle i!}\right],
 \label{EQ-22}
 \end{array}
 \end{equation}
 \begin{equation}
 \begin{array}{lll}
  V_b=U_\odot\cos l\sin b+V_\odot\sin l\sin b-W_\odot\cos b\\
  +R_0\sin l\sin b\left[\sum\limits_{i=1}^N(R-R_0)^i
  {\displaystyle\Omega_0^{(i)}\over \displaystyle i!}\right],
 \label{EQ-33}
 \end{array}
 \end{equation}
where $N$ is the order of the expansion. The quantity $\Omega_0$ is the angular velocity of rotation of the Galaxy at the solar distance $R_0$, $R$~is the distance from the
star to the axis of galactic rotation $R^2=d^2\cos^2 b-2R_0 d\cos b\cos l+R^2_0$,
the parameters $\Omega^i_0$ are the corresponding derivatives of this angular velocity, the linear velocity of rotation of the Galaxy around the galactic center in the solar neighbourhood is $V_0=|\Omega_0|R_0$. Note that according to the rectangular coordinate system we have chosen, positive rotations are rotations from axis 1 to 2~($\Omega_z$), from axis 2 to 3~($\Omega_x$), from axis 3 to 1~($\Omega_y$). In this case, the angular velocity of rotation of the Galaxy $\Omega_0$ is negative. Equations~(\ref{EQ-11})--(\ref{EQ-33}) are written accordingly. This is how they differ from the analogous equations used in the work of \cite{Bob2021}. In this paper, the value of the distance from the Sun to the Galactic center is assumed to be $R_0=8.1\pm0.1$~kpc according to the review by \cite{BB2021c}.

The system of conditional equations (\ref{EQ-11})--(\ref{EQ-33}) is solved by the least squares method (LSM) with weights of the following form:
 \begin{equation}\label{weights}
 \begin{array}{lll}
 w_r =S_0/\sqrt{S_0^2+\sigma_{V_r}^2}, \\
 w_l =S_0/\sqrt{S_0^2+\sigma_{V_l}^2}, \\
 w_b =S_0/\sqrt{S_0^2+\sigma_{V_b}^2},
 \end{array}
\end{equation}
where $\sigma_{V_r}, \sigma_{V_l}, \sigma_{V_b}$ are variances of errors of the corresponding observed velocities, $S_0$ is ``sample variance''. The value of $S_0$ is comparable to the root-mean-square residual $\sigma_0$ obtained by solving conditional equations (\ref{EQ-11})--(\ref{EQ-33}), and in this work it is assumed to be $13$~km s$^{-1}$.

The rectangular components of the spatial velocities of stars are calculated by the formulas:
\begin{equation}\label{UVW}
 \begin{array}{lll}
U=V_r \cos l\cos b-V_l\sin l-V_b\cos l\sin b,\\
V=V_r \sin l\cos b+V_l\cos l-V_b\sin l\sin b,\\
W=V_r \sin b+V_b \cos b.
\end{array}
\end{equation}
We also use the following two very important velocities: the radial velocity $V_R$, directed from center of the Galaxy to the star, and the velocity $V_{circ}$, orthogonal to $V_R$ and directed towards the rotation of the Galaxy, which are calculated using the following formulas:
 \begin{equation}
 \begin{array}{lll}
  V_{circ}= U\sin \theta+(V_0+V)\cos \theta, \\
       V_R=-U\cos \theta+(V_0+V)\sin \theta,
 \label{VRVT}
 \end{array}
 \end{equation}
where the velocities $U,$ $V,$ and $W$ are corrected for the peculiar motion of the Sun $U_\odot,$ $V_\odot,$ and $W_\odot$, the position angle $\theta$ meets the relation $\tan\theta=y/(R_0-x)$.

\subsection{Linear Ogorodnikov-Milne model} \label{ogor-milne}
In the linear Ogorodnikov-Milne model \citep{Ogor-1965}, the observed velocity of a star ${\bf V}(d)$, having a heliocentric radius vector ${\bf d}$, up to terms
of the first order of smallness $d/R_0\ll 1$ is described by the equation in vector form:
\begin{equation}
 {\bf V}(d)={\bf V}_\odot+M{\bf d}+{\bf V'},
 \label{eq-1}
 \end{equation}
where ${\bf V}_\odot(U_\odot,V_\odot,W_\odot)$ is the peculiar velocity of the Sun relative to the grouping of the stars under consideration, {$\bf V'$} is the residual velocity of a star, $M$ is a matrix (tensor) of displacements, the components of which are the partial derivatives of the velocity ${\bf u}(u_1,u_2,u_3)$ with respect to the distance ${\bf d}(d_1, d_2, d_3)$, where ${\bf u}={\bf V}(R)-{\bf V}(R_0)$, and $R$ and $R_0$ are the galactocentric distances of the star and the Sun, respectively, then
\begin{equation}
 M_{pq}={\left(\frac{\partial u_p} {\partial d_q}\right)}_\circ, \quad p,q=1,2,3,
 \label{eq-2}
 \end{equation}
where the zero means that the derivatives are taken at the point $R=R_0$. All nine elements of the $M$ matrix are determined using three components of the observed velocities~--- line-of-sight velocity $V_r,$ velocity along the galactic longitude $V_l$ and along the galactic latitude $V_b$:
 \begin{equation}
  \begin{array}{lll}
 V_r=-U_{\odot}\cos b\cos l
   -V_{\odot}\cos b\sin l-W_{\odot}\sin b\\
   +d[\cos^2 b\cos^2 l M_{11}+\cos^2 b\cos l\sin l M_{12}\\
   +\cos b\sin b \cos l  M_{13}+\cos^2 b\sin l\cos l M_{21}\\
   +\cos^2 b\sin^2 l M_{22}+\cos b\sin b\sin l M_{23}\\
   +\sin b\cos b\cos lM_{31}+\cos b\sin b\sin lM_{32}
   +\sin^2 b M_{33}], \label{eq-3}
 \end{array}
 \end{equation}
 \begin{equation}
 \begin{array}{lll}
  V_l= U_\odot\sin l-V_\odot\cos l\\
 +d [-\cos b\cos l\sin l  M_{11} -\cos b\sin^2 l M_{12}\\
 -\sin b \sin l  M_{13}+\cos b\cos^2 l M_{21}\\
 +\cos b\sin l\cos l M_{22}+\sin b\cos l  M_{23} ], \label{eq-4}
\end{array}
 \end{equation}
 \begin{equation}
 \begin{array}{lll}
 V_b=U_\odot\cos l\sin b
 +V_\odot\sin l\sin b-W_\odot\cos b\\
 +d [-\sin b\cos b\cos^2 l M_{11}\\
 -\sin b\cos b\sin l \cos l M_{12}\\
 -\sin^2 b \cos l  M_{13} -\sin b\cos b\sin l\cos l M_{21}\\
 -\sin b\cos b\sin^2 l  M_{22} -\sin^2 b\sin l  M_{23}   \\
 +\cos^2 b\cos l M_{31} +\cos^2 b\sin l M_{32}
 +\sin b\cos b  M_{33} ].
   \label{eq-5}
  \end{array}
 \end{equation}
To estimate the values of the velocities $(U,V,W)_\odot$ and the elements of the matrix $M$, the system of conditional equations (\ref{eq-3})--(\ref{eq-5}) is solved by the LSM method with weights of the form (\ref{weights}).

Matrix $M$ is divided into symmetric $M^{\scriptscriptstyle+}$ (local deformation tensor) and asymmetric $M^{\scriptscriptstyle-}$ (rotation tensor) parts:
 \begin{equation}
 \renewcommand{\arraystretch}{2.2}
  \begin{array}{lll}\displaystyle
 M_{\scriptstyle pq}^{\scriptscriptstyle+}=
 {1\over 2}\left( \frac{\partial u_{p}}{\partial d_{q}}+
 \frac{\partial u_{q}}{\partial d_{p}}\right)_\circ,  \\
 \displaystyle
 M_{\scriptstyle pq}^{\scriptscriptstyle-}=
 {1\over 2}\left(\frac{\partial u_{p}}{\partial d_{q}}-
 \frac{\partial u_{q}}{\partial d_{p}}\right)_\circ, \qquad
  p,q=1,2,3,
 \label{eq-456}
 \end{array}
 \end{equation}
where the zero means that the derivatives are taken at the point $R=R_0$. The quantities
 $M_{\scriptscriptstyle32}^{\scriptscriptstyle-},
  M_{\scriptscriptstyle13}^{\scriptscriptstyle-},
  M_{\scriptscriptstyle21}^{\scriptscriptstyle-}$
are the components of the vector of solid-body rotation of the small circumsolar neighborhood around the axes $x,y,z$, respectively. In accordance with chosen rectangular coordinate system, the positive rotations are rotations
  from axis 1 to 2~($\Omega_z$),
  from axis 2 to 3~($\Omega_x$),
  from axis 3 to 1~($\Omega_y$):
 \begin{equation}\label{Omega-0}
 M^{\scriptscriptstyle-}=\begin{pmatrix}
          0&-\Omega_z &~~\Omega_y\\
 ~~\Omega_z&         0&-\Omega_x \\
  -\Omega_y&~~\Omega_x&         0\end{pmatrix}.
 \end{equation}
The components of the rotation tensor are calculated from the elements of the matrix $M$ as follows:
\begin{equation}\label{B-xyz}
 \begin{array}{lll}
    M_{\scriptscriptstyle32}^{\scriptscriptstyle-}=
0.5(M_{\scriptscriptstyle32}-M_{\scriptscriptstyle23}),\\
    M_{\scriptscriptstyle13}^{\scriptscriptstyle-}=
0.5(M_{\scriptscriptstyle13}-M_{\scriptscriptstyle31}),\\
    M_{\scriptscriptstyle21}^{\scriptscriptstyle-}=
0.5(M_{\scriptscriptstyle12}-M_{\scriptscriptstyle21}).
\end{array}
\end{equation}
Each of the quantities
 $M_{\scriptscriptstyle12}^{\scriptscriptstyle+},
 M_{\scriptscriptstyle13}^{\scriptscriptstyle+},
 M_{\scriptscriptstyle23}^{\scriptscriptstyle+}$
describes the deformation in the corresponding plane. From the elements of the matrix $M$ they are calculated as follows:
\begin{equation}\label{A-xyz}
 \begin{array}{lll}
    M_{\scriptscriptstyle12}^{\scriptscriptstyle+}=
0.5(M_{\scriptscriptstyle12}+M_{\scriptscriptstyle21}),\\
    M_{\scriptscriptstyle13}^{\scriptscriptstyle+}=
0.5(M_{\scriptscriptstyle13}+M_{\scriptscriptstyle31}),\\
    M_{\scriptscriptstyle23}^{\scriptscriptstyle+}=
0.5(M_{\scriptscriptstyle23}+M_{\scriptscriptstyle32}).
\end{array}
\end{equation}
Diagonal components of the local deformation tensor
 $M_{\scriptscriptstyle11}^{\scriptscriptstyle+},
  M_{\scriptscriptstyle22}^{\scriptscriptstyle+},
  M_{\scriptscriptstyle33}^{\scriptscriptstyle+},$
coincide with the corresponding diagonal elements of the matrix $M$. They describe a general local contraction or expansion of the entire stellar system (divergence). In particular, it is interesting to evaluate the volumetric expansion/contraction effect:
 \begin{equation}
 K_{xyz}= (M_{\scriptscriptstyle11}+
           M_{\scriptscriptstyle22}+
           M_{\scriptscriptstyle33})/3.
 \label{K-XYZ}
 \end{equation}
The Ogorodnikov-Milne model has been successfully used by many authors to study stellar motions in different planes \citep{Clube1972,Mont1977,Miyamoto1998,Zhu2000,BobKhov2011,Velichko2020}. In this paper, we are mainly interested in the vertical movements of stars.

\subsection{Formation of residual velocities} \label{residual}
We need the residual velocities in order to discard the large discrepancies when determining the parameters of the rotation of the Galaxy. In addition, we need residual velocities to exclude from the observed velocities the influence of the differential rotation of the Galaxy in the analysis of effects not related to the rotation of the Galaxy.

The residual velocities of the Cepheids are calculated taking into account the peculiar velocity of the Sun, $U_\odot, V_\odot$ and $W_\odot$, and the differential rotation of the Galaxy, in the following form:
\begin{equation}
 \begin{array}{lll}
 V_r=V^*_r
 -[-U_\odot\cos b\cos l-V_\odot\cos b\sin l-W_\odot\sin b\\
 -R_0(R-R_0)\sin l\cos b\Omega^\prime_0\\
 -0.5R_0(R-R_0)^2\sin l\cos b\Omega^{\prime\prime}_0
 -\ldots],
 \label{EQU-1}
 \end{array}
 \end{equation}
 \begin{equation}
 \begin{array}{lll}
 V_l=V^*_l
 -[U_\odot\sin l-V_\odot\cos l-d\Omega_0\cos b\\
 -(R-R_0)(R_0\cos l-d\cos b)\Omega^\prime_0\\
 -0.5(R-R_0)^2(R_0\cos l-d\cos b)\Omega^{\prime\prime}_0
 -\ldots],
 \label{EQU-2}
 \end{array}
 \end{equation}
  \begin{equation}
 \begin{array}{lll}
 V_b=V^*_b
 +[U_\odot\cos l\sin b + V_\odot\sin l\sin b-W_\odot\cos b\\
 +R_0(R-R_0)\sin l\sin b\Omega^\prime_0\\
 +0.5R_0(R-R_0)^2\sin l\sin b\Omega^{\prime\prime}_0
 +\ldots],
 \label{EQU-3}
 \end{array}
 \end{equation}
where the velocities $V^*_r,V^*_l,V^*_b,$ on the right-hand sides of the equations are the initial velocities, and on the left-hand sides of the equations there are the corrected velocities $V_r,V_l,V_b,$ using which you can calculate the residual velocities $U,V,W$ by the formulas (\ref{UVW}).

 \begin{figure*} {\begin{center}
   \includegraphics[width=0.85\textwidth]{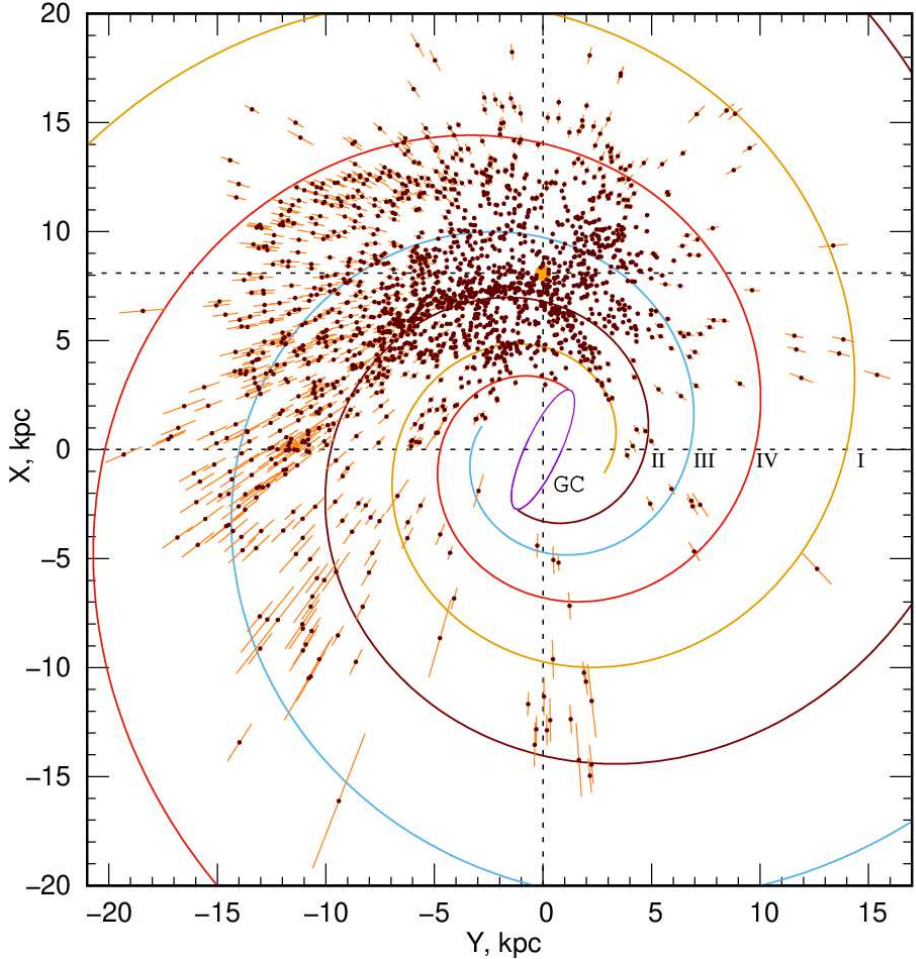}
 \caption{
The distribution of Cepheids in the projection on the galactic plane $XY,$ the position of the Sun is marked with an orange circle $(X,Y)=(8.1,0)$~kpc, see text.
 }
 \label{f-XY-1}
 \end{center} }
 \end{figure*}

 \section{DATA} \label{data}
In this work, we use data on classical Cepheids from the works of~\cite{Skowron19a}. These Cepheids were observed as part of the fourth stage of the OGLE program (Optical Gravitational Lensing Experiment, \cite{Udalski2015}). The~\cite{Skowron19a} catalog contains estimates of distance, age, pulsation period, and photometric data for 2431 Cepheids. Their apparent magnitudes lie in the range of $11^m<I<18^m$, so there is a small deficit of bright and well-studied Cepheids known from earlier observations.

The heliocentric distances up to 2214 Cepheids, $d$, were calculated by~\cite{Skowron19a} based on the period--luminosity relationship. The specific relationship they used was refined by \cite{Wang2018a} from the light curves of Cepheids in the mid-infrared range, where the interstellar absorption is significantly less than in the optical one. The estimates of the age of Cepheids in the work of~\cite{Skowron19a} were made according to the method developed by \cite{Anderson2016}, which took into account the periods of axial rotation of stars and the metallicity indices.

We identified 2214 Cepheids with the distances from~\cite{Skowron19a} with the Gaia\,EDR3 catalog. As a result, a sample of about 2000 stars with proper motions, heliocentric distances, and age estimates was obtained. For about 750 stars from the resulting list, there are line-of-sight velocities from the Gaia\,DR2 catalog.

As a result, we have a sample of stars with very remarkable characteristics. According to estimates~\cite{Skowron19a}, the distances to Cepheids in their catalog are determined with an average relative error of about 5\%. This high accuracy was achieved by these authors using infrared data from Spitzer~\citep{Benjamin2003,Churchwell2009} and WISE~\citep{Wright2010,Mainzer2011}. The distances to the Cepheids were calculated based on the period-luminosity relation obtained by \cite{Wang2018a} for the infrared range. Interstellar extinction was estimated using extinction maps by \cite{Bovy2016}.

Random errors of estimates of the line-of-sight velocities of Cepheids from the Gaia\,DR2 catalog are in average 5~km s$^{-1}$ (see, for example, fig.1a in \cite{Bob2021}). Finally, compared with the Gaia\,DR2, the Gaia\,EDR3 catalog has about two times better \citep{Brown21} accuracy in determining the stellar proper motions. Taking the value of the error in determining the proper motion along one of the coordinates equal to 0.03~mas yr$^{-1}$, we can estimate that the error in the stellar tangential velocity will be about 5~km s$^{-1}$ at a distance from the Sun of more than 15~kpc. We can conclude that we are able to analyze high-precision spatial velocities of Cepheids in a wide range of distances $R$ (up to $R\sim23$~kpc).

\section{RESULTS} \label{results}
\subsection{Restrictions}
Conditional equations of the form (\ref{EQ-11})--(\ref{EQ-33}) or (\ref{eq-1})--(\ref{eq-3}) are usually solved together using all three components $V_r$, $V_l$ and $V_b$. True, line-of-sight velocities are not known for all Cepheids. When the star has all three velocities $V_r$, $V_l$ and $V_b$, then we apply the following restrictions:
 \begin{equation}
 \begin{array}{rcl}
   |U|<80~\hbox{km s$^{-1}$},\\
   |V|<80~\hbox{km s$^{-1}$},\\
   |W|<60~\hbox{km s$^{-1}$},
     \label{cut-Vr}
 \end{array}
 \end{equation}
where the velocities $U,V,W$ are residual, i.e. freed from the influence of galactic rotation. When for a star there are only two velocities $V_l$ and $V_b$, then we apply the following restrictions:
 \begin{equation}
 \begin{array}{rcl}
   |U|<120~\hbox{km s$^{-1}$},\\
   |V|<120~\hbox{km s$^{-1}$},\\
   |W|<~60~\hbox{km s$^{-1}$}.
  \label{cut-Vlb}
 \end{array}
 \end{equation}
We do not use Cepheids, which have a large residual velocity dispersion. These are Cepheids located near the galactic center in the region of radius $R=3$~kpc, where the central bar strongly distorts the kinematics of stars. We do not use Cepheids, designated in the~\cite{Skowron19a} catalog as OGLE\,BLG-CEP-XXX~--- they belong to the bulge, sometimes located at very large heliocentric distances (up to $d=25$~kpc), far behind the center of the Galaxy.

The distribution of 1847 Cepheids selected using all spatial and kinematic constraints in projection onto the galactic plane $XY$ are given in Fig.~\ref{f-XY-1}. For the figure, a coordinate system was used, where $X$ is directed from the center of the Galaxy, the direction of the $Y$ axis coincides with the direction of the previously described $Y$ axis. Note that one Cepheid with distances of $Y\sim-22$~kpc was left behind the frame. The figure shows a four-arm spiral pattern with a pitch angle $-13^\circ$ \citep{BobBajk14}, the following segments of the spiral arms are numbered in Roman numerals: I~--- Scutum-Centarius arm, II~--- Carina-Sagittarius arm, III~--- Perseus arm, and IV~--- Outer arm, the central galactic bar is shown schematically. The concentration of Cepheids towards the Carina-Sagittarius arm is clearly visible.

 \begin{figure*} {\begin{center}
   \includegraphics[width=0.9\textwidth]{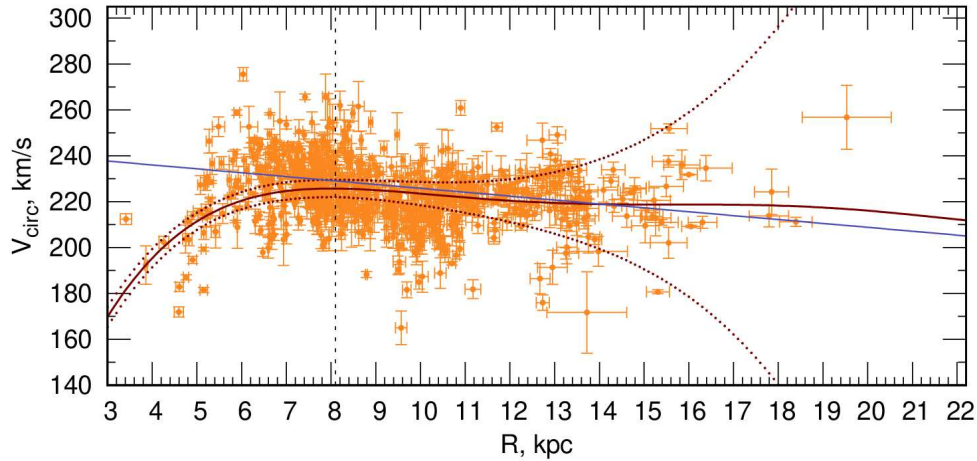}
 \caption{
Circular velocities $V_ {circ}$ of a sample of Cepheids with proper motions and radial velocities. The position of the Sun is marked with a vertical dashed line at distance $R_0=8.1$~kpc, a galactic rotation curve corresponding to the solution~(\ref{solution-24}) is given, with indicating the boundaries of the confidence intervals corresponding to the level of errors $1\sigma$, the blue line shows the rotation curve from the work of \citep{Eilers2019}.
 }
 \label{f1-rotcurve}
 \end{center} }
 \end{figure*}

 \subsection{Galactic rotation parameters}
Table~\ref{t1} gives the parameters of the Galaxy rotation, found as a result of the LSM solution of a system of conditional equations~(\ref{EQ-11})--(\ref{EQ-33}). Restrictions~(\ref{cut-Vr})--(\ref{cut-Vlb}) were used. The calculations were carried out for five samples with different boundaries of the $r$ distance.

As can be seen from Table~\ref{t1}, the values of the components of the angular velocity of rotation change depending on the radius of the sample. In particular, the value of the linear rotation velocity $V_0$ changes slightly from 227~km s$^{-1}$ to 236~km s$^{-1}$. In each case, a rotation curve is perfect for describing rotation in the appropriate range of distances.

The nonlinear model of the Galaxy's rotation can be applied at any sample radius, having a sufficient number of derivatives in the expansion of the angular rotation velocity. However, the model includes a number of local parameters (to determine which it is sufficient to use a small neighborhood $R\rightarrow R_0$). These are the velocities $(U,V,W)_\odot$ and $\Omega_0$ (and hence $V_0$). It is better to estimate these values on a sample that satisfies the completeness condition in order to obtain unbiased estimates.

The question of the completeness boundary of Cepheids from the catalog of~\cite{Skowron19a} was considered in the work of \cite{BB2021a}. It turned out that, depending on the age, these Cepheids are not quite equally distributed. In particular, it was shown that for a sample of Cepheids older than 130 Ma, the completeness limit is 6.5~kpc. Middle-aged Cepheids, from 80 to 130 Myr, almost all of them are concentrated in the section of the Carina-Sagittarius spiral arm. Finally, for Cepheids younger than 80 Myr, the completeness limit is 4.5~kpc. As a result, in the work of \cite{BB2021a}, the value $r=5$~kpc was adopted, which corresponds to the boundary where the sample of Cepheids under consideration satisfies the completeness condition. However, we believe that we have a sufficient number of Cepheids to reliably determine the parameters of galactic rotation in a fairly wide range of distances $R$.

The Galactic rotation curve constructed from the data from the penultimate column of Table~\ref{t1} ($d\leq10$~kpc) sharply turns downward at $R\sim17$~kpc. And the rotation curve constructed according to the data from the first column of Table~\ref{t1} turns down at $R\sim19$~kpc. But in our sample there are Cepheids located at larger distances. For the study of warp kinematics, the most distant stars are of the greatest interest.

We also obtained a solution using almost the entire sample of Cepheids. In it, in particular, the following values of the velocity of the Sun were found $(U_\odot,V_\odot,W_\odot)= (9.39,15.96,6.88)\pm(0.40,0.55,0.30)~\hbox{km s$^{-1}$},$ and also:
 \begin{equation}
 \label{solution-24}
 \begin{array}{cll}
       \Omega_0=-27.87\pm0.09~\hbox{km s$^{-1}$ kpc$^{-1}$},\\
   \Omega^{'}_0=~3.461\pm0.033~\hbox{km s$^{-1}$ kpc$^{-2}$},\\
  \Omega^{''}_0=-0.651\pm0.024~\hbox{km s$^{-1}$ kpc$^{-3}$},\\
 \Omega^{'''}_0=~0.105\pm0.011~\hbox{km s$^{-1}$ kpc$^{-4}$},\\
  \Omega^{IV}_0=-0.009\pm0.002~\hbox{km s$^{-1}$ kpc$^{-5}$},\\
 \end{array}
 \end{equation}
where the unit weight error is $\sigma_0=12.62$~km s$^{-1}$, and the linear velocity of rotation of the circumsolar neighborhood is $V_0=225.7\pm2.9$~km s$^{-1}$ for the accepted value $R_0=8.1\pm0.1$~kpc. Here, 1848 Cepheids from the range of galactocentric distances $R:3-24$~kpc were used.

The galactic rotation curve corresponding to the solution~(\ref{solution-24}) is shown in Fig.~\ref{f1-rotcurve}. It is interesting to note that this curve begins to sharply tend upward only at $R\sim26$~kpc. The circular velocities in this figure are given for 732 stars with known proper motions and line-of-sight velocities. We can conclude that the rotation curve corresponding to the solution~(\ref{solution-24}) is flatter at the largest distances $R$ than the curve obtained by us earlier \citep{Bob2021} by a similar method from the~\cite{Mroz19} catalog.

Fig.~\ref{f1-rotcurve} also shows the rotation curve from the work of \cite{Eilers2019}. This is one of the modern rotation curves of the Galaxy, constructed for a wide range of galactocentric distances. To build it, these authors used highly accurate data on over 23\,000 luminous red giant stars, which are distributed over a distance interval of $R:5-25$~kpc. These authors found $V_0=229.0\pm0.2$~km s$^{-1}$ (for adopted $R_0=8.122$~kpc) with a derivative of $-1.7\pm0.1$ km s$^{-1}$ kpc$^{-1}$. As can be seen from this figure, there is good agreement between our curve and the curve from the work of \cite{Eilers2019} both in the solar region, at $R=R_0$, and far in the outer region of the Galaxy.

We decided further to use the solution~(\ref{solution-24}) to form the residual velocities, since the corresponding galactic rotation curve covers the largest range of distances $R:3-24$~kpc.

{\begin{table*}
\caption[]{\small\baselineskip=1.0ex\protect
Galaxy rotation parameters found based on a nonlinear model }
\begin{center}\small\protect
\label{t1}
\begin{tabular}{|l|r|r|r|r|r|}\hline
 Par. & $d\leq16$~kpc & $d\leq14$~kpc & $d\leq12$~kpc & $d\leq10$~kpc & $d\leq5$~kpc\\\hline

 $N_\star$  &         1766 &         1707 &         1598 &         1480 &   835\\
 $\sigma_0$ &        12.35 &        12.24 &        12.15 &        12.03 & 11.69\\ &&&&& \\
 $U_\odot$  &$ 9.26\pm0.40$&$ 9.17\pm0.40$&$ 8.85\pm0.40$&$ 8.83\pm0.40$&$ 9.56\pm0.46$ \\
 $V_\odot$  &$15.77\pm0.58$&$15.38\pm0.59$&$15.03\pm0.59$&$14.84\pm0.59$&$13.96\pm0.71$ \\
 $W_\odot$  &$ 6.51\pm0.30$&$ 6.42\pm0.30$&$ 6.38\pm0.31$&$ 6.35\pm0.32$&$ 6.85\pm0.41$ \\          &&&&& \\
 $\Omega_0$ &$-27.97\pm0.09$ & $-28.03\pm0.09$ & $-28.28\pm0.10$ & $-28.48\pm0.10$ & $-29.18\pm0.18$ \\
 $\Omega^\prime_0$ &$  3.478\pm0.035$ & $  3.504\pm0.036$ & $  3.576\pm0.039$ & $  3.714\pm0.043$& $  4.231\pm0.075$ \\
 $ \Omega^{\prime\prime}_0$ &$ -0.643\pm0.026$ & $ -0.625\pm0.027$ & $ -0.613\pm0.027$ & $ -0.621\pm0.028$& $ -0.782\pm0.070$  \\
 $\Omega^{\prime\prime\prime}_0$ &$  0.093\pm0.016$ & $  0.073\pm0.018$ & $  0.043\pm0.020$ & $ -0.023\pm0.024$& $ -0.306\pm0.049$  \\
 $ \Omega^{IV}_0$ &$ -0.004\pm0.004$ & $  0.001\pm0.005$ & $  0.011\pm0.006$ & $  0.040\pm0.008$& $ 0.250\pm0.060$  \\
           &&&&& \\
 $|V_0|$ & $226.6\pm2.9$ & $227.1\pm2.9$ & $229.0\pm2.9$ & $230.7\pm3.0$ & $236.3\pm3.3$ \\\hline

\end{tabular}
\end{center}
{\small\baselineskip=1.0ex\protect
The velocities $(U,V,W)_\odot,$ $V_0$ and $\sigma_0$ are given in km s$^{-1}$,
  $\Omega_0$ in km s$^{-1}$ kpc$^{-1}$,
  $\Omega^\prime_0$ in km s$^{-1}$ kpc$^{-2}$,
  $\Omega^{\prime\prime}_0$ in km s$^{-1}$ kpc$^{-3}, $
  $\Omega^{\prime\prime\prime}_0$ in km s$^{-1}$ kpc$^{-4}$ and
  $\Omega^{IV}_0$ in km s$^{-1}$ kpc$^{-5}$.
   }
\end{table*}
}
{\begin{table*}                                                
\caption[]{\small\baselineskip=1.0ex\protect
Kinematic parameters of the complete linear Ogorodnikov-Milne model }
\begin{center}\small
\label{t2}
\begin{tabular}{|l|r|r|r|r|r|}\hline
  Par. & All & $d\leq16$~kpc & $d\leq14$~kpc & $d\leq12$~kpc & $d\leq2$~kpc   \\\hline

 $N_\star$ &  1866 &  1773 &  1711 &  1601 &   197 \\
 $\sigma_0$& 13.96 & 13.24 & 12.95 & 12.68 & 10.14 \\
           &       &&&&  \\
 $U_\odot$ & $ 9.31\pm0.45$ & $ 9.33\pm0.43$ & $ 9.23\pm0.42$ & $ 9.31\pm0.42$& $ 8.37\pm1.01$\\
 $V_\odot$ & $15.01\pm0.46$ & $15.50\pm0.44$ & $15.68\pm0.43$ & $16.08\pm0.43$& $ 13.89\pm0.90$\\
 $W_\odot$ & $ 5.44\pm0.40$ & $ 5.64\pm0.39$ & $ 5.67\pm0.38$ & $ 5.62\pm0.38$& $ 5.75\pm0.84$\\
           &                &&&&  \\
 $M_{11}$ & $ 0.40\pm0.14$ & $ 0.56\pm0.14$ & $ 0.73\pm0.15$ & $ 0.66\pm0.15$ & $ -0.33\pm1.01$ \\
 $M_{12}$ & $ 0.15\pm0.09$ & $ 0.16\pm0.09$ & $ 0.32\pm0.09$ & $ 0.33\pm0.09$ & $ -0.03\pm0.76$ \\
 $M_{13}$ & $-1.52\pm1.20$ & $-3.89\pm1.30$ & $-8.39\pm1.39$ & $-9.65\pm1.57$ & $ -36.30\pm8.42$ \\

 $M_{21}$ & $ 0.72\pm0.13$ & $ 0.32\pm0.12$ & $ 0.11\pm0.13$ & $-0.24\pm0.14$ & $ 3.67\pm0.90$ \\
 $M_{22}$ & $ 0.17\pm0.13$ & $ 0.22\pm0.13$ & $ 0.14\pm0.13$ & $-0.03\pm0.13$ & $ -2.20\pm0.82$ \\
 $M_{23}$ & $ 4.72\pm1.44$ & $ 3.04\pm1.67$ & $ 3.32\pm1.67$ & $ 3.82\pm1.77$ & $ 8.97\pm8.30$ \\

 $M_{31}$ & $-0.49\pm0.08$  & $-0.48\pm0.10$ & $-0.53\pm0.10$ & $-0.63\pm0.11$ & $ -0.35\pm0.86$ \\
 $M_{32}$ & $ 0.06\pm0.08$  & $-0.02\pm0.08$ & $-0.09\pm0.08$ & $-0.07\pm0.08$ & $ 0.70\pm0.73$ \\
 $M_{33}$ & $ 1.55\pm1.02$  & $ 1.27\pm1.20$ & $ 2.12\pm1.26$ & $ 2.12\pm1.43$ & $ -4.75\pm8.30$ \\
          &                 &&&&  \\
$K_{xyz}$ & $0.71\pm0.35$ & $0.68\pm0.40$ & $1.00\pm0.43$ & $0.92\pm0.48$ & $-2.42\pm2.80$ \\\hline
\end{tabular}
\end{center}
{\small\baselineskip=1.0ex\protect
The velocities $(U,V,W)_\odot$ and $\sigma_0$ are given in km s$^{-1}$, $M_{p,q}, p,q=1,2,3$ and $K_{xyz}$ in km s$^{-1}$ kpc$^{-1}$.
}
\end{table*}
}

 \begin{figure*} {\begin{center}
 \includegraphics[width=0.97\textwidth]{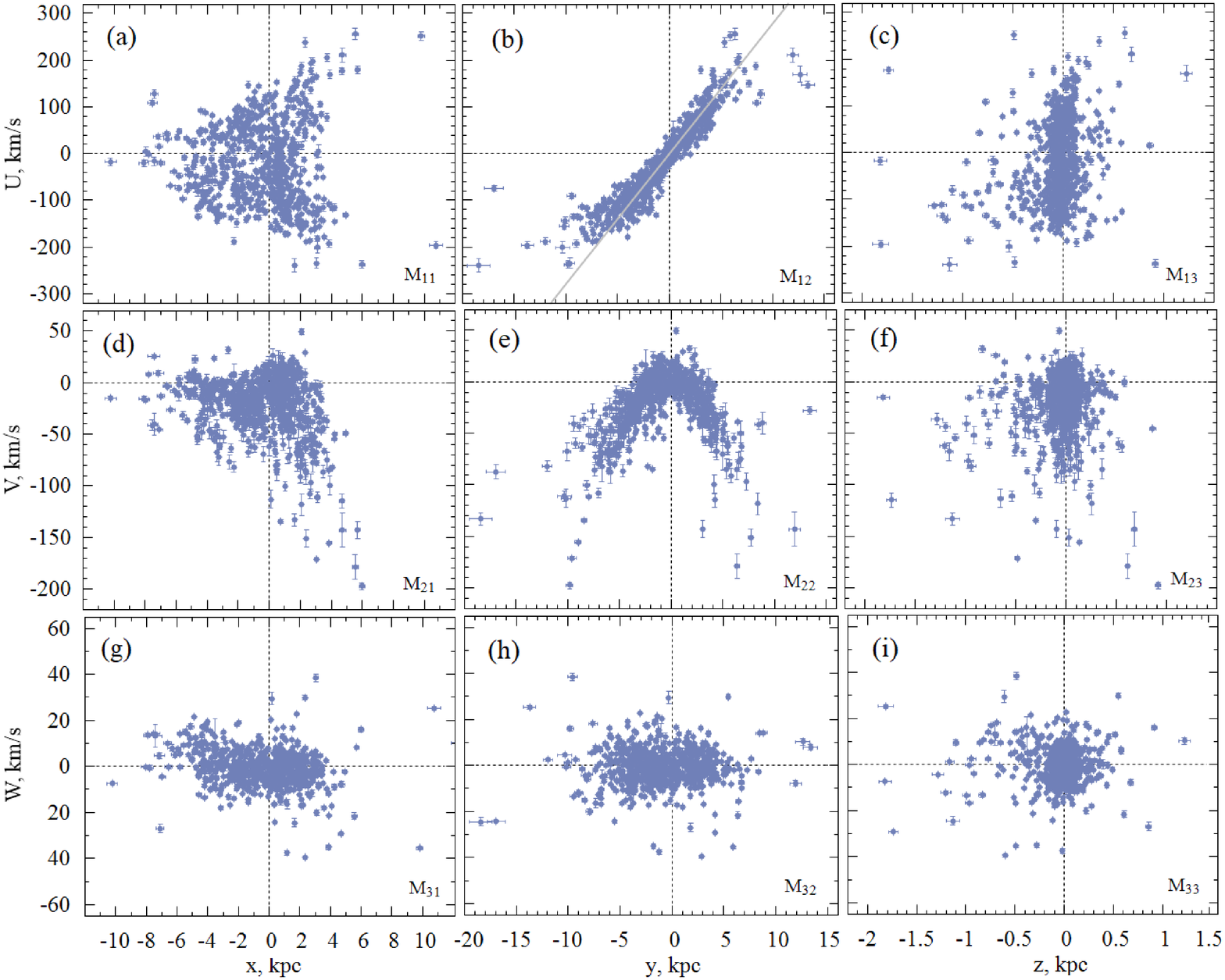}
 \caption{
Velocities of Cepheids $U,V,W$ versus heliocentric rectangular coordinates $x,y,z$. On each panel, the corresponding designations of the deformation matrix $M_{p,q}$ are given, on panel (b) the gray line shows the dependence $U=M_{12}\cdot y$, where
$M_{12}=27.87\pm0.09$~km s$^{-1}$ kpc$^{-1}$ corresponds to the solution (\ref{solution-24}).
 }
 \label{f-3}
 \end{center} }
 \end{figure*}

Fig.~\ref{f-3} shows nine dependences of the velocities $U,V,W$ on the corresponding coordinates $x,y,z$. Fig.~\ref{f-3}\,b shows the dependence $U=M_{12}\cdot y$, where $M_{12}$ corresponds to the angular velocity of rotation of the Galaxy around the axis $z$ (see (\ref{Omega-0})), $M_{12}=\partial U/\partial y=-\Omega_z=27.87\pm0.09$~km s$^{-1}$ kpc$^{-1}$. It can be seen that the points on this graph follow this linear dependence well in a wide range of $-10<y<10$~kpc. But on the other five panels with velocities $U$ and $V$, the nonlinear character of the velocity distribution is clearly visible.

 \subsection{Residual velocity analysis}
Table~\ref{t2} shows the parameters of the Ogorodnikov-Milne linear model, found as a result of the LSM solution of a system of conditional equations~(\ref{eq-3})--(\ref{eq-5}). In this case, the velocities $V_r,V_l,V_b$ were corrected for the differential rotation of the Galaxy using the relations~(\ref{EQU-1})--(\ref{EQU-3}), as well as the angular velocity parameters~(\ref{solution-24}). Restrictions~(\ref{cut-Vr})--(\ref{cut-Vlb}) were used.

At the top of the table, the number of used stars $N_\star$, the error of the unit of weight $\sigma_0,$ found as a result of the LSM solution of a system of conditional equations~(\ref{eq-3})--(\ref{eq-5}), the solar peculiar velocity components $(U,V,W)_\odot$ and nine elements of the $M$ matrix (\ref{eq-2}) are given. The calculations were performed for five samples with different boundaries of the heliocentric distance $r$.

It is well known \citep{Ogor-1965} that in studying the rotation of the Galaxy the linear Ogorodnikov-Milne model is applicable up to distances $d\leq2$~kpc. This can be clearly seen from the behavior of the velocities $U$ and $V$ in Fig.~\ref{f-3}. Therefore, the last column of Table~\ref{t2} contains the solution obtained under this constraint. On the other hand, we hope that the analysis of the residual velocities of stars can be carried out at any sample radius.

We also use the simplest pure rotation model. The conditional equations are as follows:
 \begin{equation}
  \begin{array}{lll}
  V_l= U_\odot\sin l-V_\odot\cos l\\
 +d[-\cos l\sin b M_{\scriptscriptstyle32}^{\scriptscriptstyle-}
    -\sin l\sin b M_{\scriptscriptstyle13}^{\scriptscriptstyle-}
    +\cos b       M_{\scriptscriptstyle21}^{\scriptscriptstyle-}],\label{eq-44}
\end{array}
 \end{equation}
 \begin{equation}
 \begin{array}{lll}
 V_b=U_\odot\cos l\sin b+V_\odot\sin l\sin b-W_\odot\cos b\\
 +d [\sin l M_{\scriptscriptstyle32}^{\scriptscriptstyle-}
    -\cos l M_{\scriptscriptstyle13}^{\scriptscriptstyle-}  ].
   \label{eq-55}
  \end{array}
 \end{equation}
The velocities $V_l$ and $V_b$ were corrected for the differential rotation of the Galaxy using the relations~(\ref{EQU-2})--(\ref{EQU-3}) and constraints (\ref{cut-Vr})--(\ref{cut-Vlb}). The rotation curve parameters were taken according to the solution~(\ref{solution-24}).

Table~\ref{t3} gives the found velocities $(U,V,W)_\odot$ and three angular velocities of rotation
 $M_{\scriptscriptstyle32}^{\scriptscriptstyle-}=\Omega_x$
 $M_{\scriptscriptstyle13}^{\scriptscriptstyle-}=\Omega_y$
 $M_{\scriptscriptstyle21}^{\scriptscriptstyle-}=\Omega_z$.

 \begin{figure*} {\begin{center}
  \includegraphics[width=0.97\textwidth]{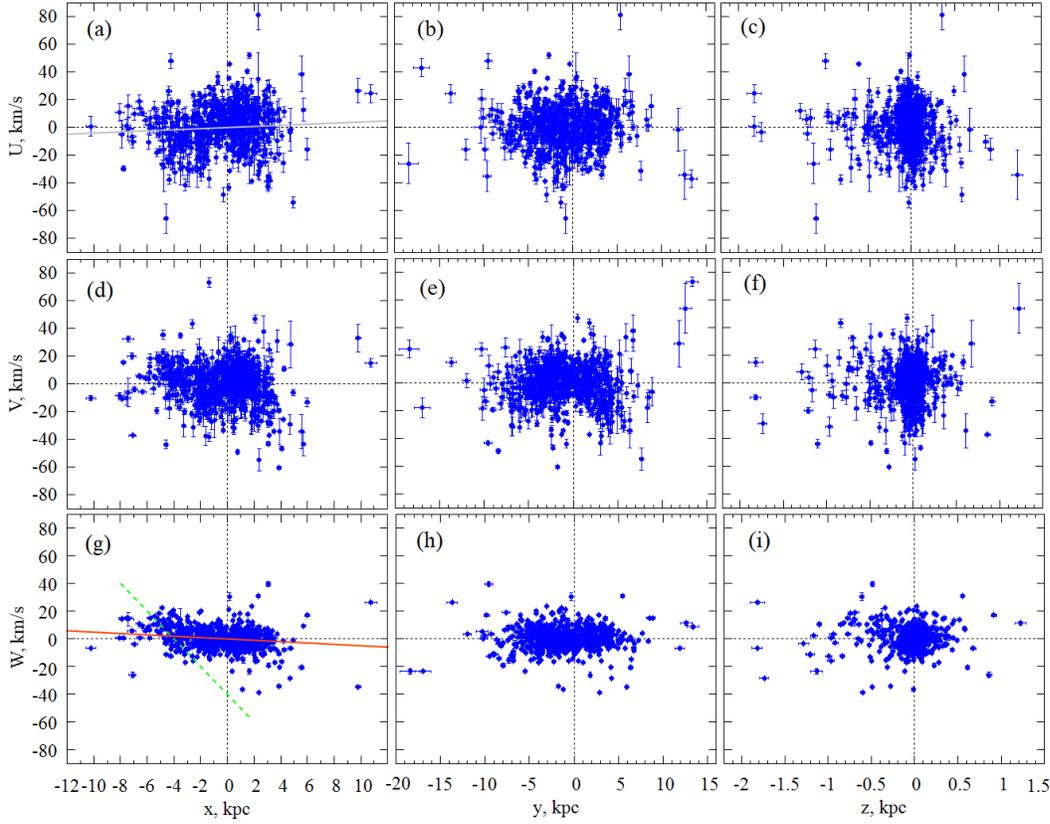}
 \caption{
Velocities of the Cepheids $U,V,W$ (corrected for the rotation of the Galaxy with parameters (\ref{solution-24})) versus the heliocentric rectangular coordinates $x,y,z$. In panel (b), the gray line shows the dependence
$U=M_{11}\cdot x$, where $M_{11}=0.40\pm0.14$~km s$^{-1}$ kpc$^{-1}$, the red line on panel (g) shows the dependence $W=M_{31}\cdot x$, where $M_{31}=-0.49\pm0.08$~km s$^{-1}$ kpc$^{-1}$.}
 \label{f-4}
 \end{center} }
 \end{figure*}

Fig.~\ref{f-4} shows nine dependences of the residual velocities $U,V,W$ on the corresponding coordinates $x,y,z$. In panel~(b), the gray line shows the dependence
$U=M_{12}\cdot y$, where $M_{12}=0.40\pm0.14$~km s$^{-1}$ kpc$^{-1}$, the red line on panel (g) shows the dependence $W=M_{31}\cdot x$, where $M_{31}=-0.49\pm0.08$~km s$^{-1}$ kpc$^{-1}$.

Note that the velocities $W$ do not depend on the rotation of the Galaxy. It can be seen that in Fig.~\ref{f-4} they are the same as in Fig.~\ref{f-3}. Moreover, in Fig.~\ref{f-3}g,h,i there is a slightly larger vertical scale in comparison with Fig.~\ref{f-4}g,h,i.

We believe that the rotation around the $y$ ($M_{31}$) axis that we have found is closely related to indirect rather than direct influence of some process that perturbs the vertical velocities of stars. At large distances from the Sun, and at large distances from the center of the Galaxy,  there are known processes that perturb the vertical positions and speeds of stars. Such processes include large-scale Warp of the galactic disk, which affects the vertical positions of stars, and its deformation precession, which affects the vertical velocities of stars. Recently, large-scale perturbations of the vertical velocities of stars have also been discovered, which are associated either with the proper oscillations of the disk or with the consequences of the fall of a massive impactor (for example, a dwarf satellite galaxy) onto the disk. Vertical asymmetries have been found in both positions
and velocities, and associated with both the Galactic warp and with bending or breathing modes in the disc \citep{Widrow2012,Bennett2019,Carrillo2019,Cheng2020,Chrobakova2022,McMillan2022}.

For example~\cite{Poggio2018}, shows that the vertical velocities of distant stars begin to increase strongly when $R$ is greater, approximately 12--14~kpc. Thus, estimates of the deformation precession rate  turn out to be of the order of 10~km s$^{-1}$ kpc$^{-1}$ \citep{Poggio2020}.

We use a linear method of analysis. In this case, we use both nearby stars, with unperturbed vertical velocities, as well as stars far from the Sun, whose vertical velocities are perturbed. Therefore, we say that our estimates of the gradients reflect the indirect influence of some process that perturbs the vertical velocities of Cepheids.

For example, the green dotted line in Fig.~\ref{f-4}g shows a linear relationship $W=-10x-40$, where the coefficient, equal to 10~km s$^{-1}$ kpc$^{- 1}$, may reflect the effect of deformation precession. We took the value of this coefficient in accordance with the definition of \cite{Poggio2020}. Fig.~\ref{f-3}g shows even better that for $x<-4$~kpc the horizontal distribution of velocities $W$ is replaced by a rather sharp rise.

 \section{DISCUSSION} \label{discussion}
 \subsection{Peculiar velocity of the Sun} \label{local}
The velocities $(U,V,W)_\odot$ are the group velocity of the considered sample of Cepheids, taken with the opposite sign. These velocities contain the peculiar motion of the Sun relative to the Local Standard of Rest, disturbances from the spiral density wave (for relatively young objects), and the influence on the velocity $V_\odot$ of the so-called asymmetric drift (lagging behind the circular velocity of rotation with the age of the sample).

Asymmetric drift \cite{Stromberg1946} manifests itself in the fact that the velocity in the direction of galactic rotation, V, systematically increases with the square of the stellar velocity dispersion. The value of the correction for asymmetric drift, $V_{asym}$, strongly depends on the age of the sample stars. For example, according to \cite{Bensby2003} $V_{asym}$ is: $-15$, $-46$ and $-220$~km s$^{-1}$ for thin disc, thick disc and halo, respectively.

The components of the solar peculiar velocity relative to the Local Standard of Rest are determined rather carefully in the work of \cite{Schonrich2010}, and are $(U,V,W)_\odot=(11.1,12.2,7.3)\pm(0.7,0.5,0.4)$~km s$^{-1}$.

We note the work \cite{Tian2015}, in which a kinematic analysis of about 200\,000 relatively close to Sun FGK-type main-sequence stars selected from LAMOST \citep{Cui2012,Deng2012,Zhao2012}. These authors have found values $(U,V,W)_\odot=(9.58,10.52,7.01)\pm(2.39,1.96,1.67)$~km s$^{-1}$, close to~\cite{Schonrich2010}.

\cite{Huang2015} analyzed FGK dwarfs from the LAMOST catalog
using their line-of-sight velocities and proper motions. The stars were taken from the circumsolar region with a radius of about 600~pc. In particular, these authors found $(U,V,W)_\odot=(7.01,10.13,4.95)\pm(0.20,0.12,0.09)$~km s$^{-1}.$

It is interesting to note the work \cite{Coskunoglu2011}, where a sample of 18\,026 high-probability thin-disc dwarfs within 600~pc of the Sun from the RAVE \citep{Steinmetz2006} spectroscopic survey was analyzed. The RAVE survey is distinguished by high-precision line-of-sight velocities of stars, which are determined with average errors of about 2~km s$^{-1}$. The proper motions of the stars here were taken from various catalogs. These authors found $(U,V,W)_\odot=(8.50,13.38,6.49)\pm(0.29,0.43,0.26)$~km s$^{-1}.$ The velocity $(U,V,W)_\odot$ values we found are in very good agreement with the results of \cite{Schonrich2010} and \cite{Coskunoglu2011}.

In the work of~\cite{Mroz19}, when analyzing about 770 Cepheids from the list of~\cite{Skowron19a} with known line-of-sight velocities, the values $(U,V,W)_\odot=(10.1,12.3,7.3)\pm(1.0,2.1,0.7)$~km s$^{-1}$ (Model~2 with prior on $R_0$).

In the work of \cite{Bob2021}, when analyzing about 800 Cepheids from the list of~\cite{Skowron19a} with known line-of-sight velocities, were found the values $(U,V,W)_\odot=(10.1,13.6,7.0)\pm(0.5,0.6,0.4)$~km s$^{-1}$ on the basis of the nonlinear model of galactic rotation.

The values of the velocities $(U,V,W)_\odot$ found by us are in good agreement with these estimates. As you can see from Table~\ref{t1}, Table~\ref{t2} and Table~\ref{t3}, all three components of this velocity are stable under different constraints on the distance $r$. Note the values $(U,V,W)_\odot=(9.56,13.96,6.85)\pm(0.46,0.71,0.41)$~km s$^{-1}$ from the last column of Table~\ref{t1}, obtained from the sample Cepheids satisfying the completeness condition, as well as $(U,V,W)_\odot=(9.02,14.36,6.09)\pm(0.53,0.60,0.40)$~km s$^{-1}$ from the last column of Table~\ref{t3} obtained under a similar condition using only proper motions from the Gaia\,EDR3 catalog. Cepheids are relatively young stars (the average age of the sample is about 110 Ma), therefore, they do not have a noticeable shift in $\Delta V_\odot$ due to asymmetric drift compared to the result, for example, of \cite{Schonrich2010}.

{\begin{table*}                                                
\caption[]{\small\baselineskip=1.0ex\protect
 Kinematic parameters of the rotation model found only from the proper motions of the Cepheids }
\begin{center}\small
\label{t3}
\begin{tabular}{|l|r|r|r|r|r|}\hline
  Par. & All & $d\leq16$~kpc & $d\leq14$~kpc & $d\leq12$~kpc & $d\leq5$~kpc \\\hline

 $N_\star$ &  1823 &  1741 &  1688 &  1600 &   837\\
 $\sigma_0$& 12.47 & 11.98 & 11.79 & 11.59 & 11.17\\
           &&&&&  \\
 $U_\odot$ &$ 8.84\pm0.45$&$ 8.38\pm0.42$&$ 8.31\pm0.44$&$ 7.96\pm0.44$ &$ 9.02\pm0.53$\\
 $V_\odot$ &$13.82\pm0.51$&$14.31\pm0.50$&$14.50\pm0.49$&$14.89\pm0.49$ &$14.36\pm0.60$\\
 $W_\odot$ &$ 5.64\pm0.35$&$ 5.77\pm0.34$&$ 5.79\pm0.34$&$ 5.73\pm0.34$ &$ 6.09\pm0.40$\\
           &&&&&  \\
$M_{\scriptscriptstyle32}^{\scriptscriptstyle-}=\Omega_{x}$ &
  $ 0.02\pm0.06$ &$-0.04\pm0.06$ &$-0.05\pm0.06$ &$-0.02\pm0.07$ &$ 0.04\pm0.16$ \\
$M_{\scriptscriptstyle13}^{\scriptscriptstyle-}=\Omega_{y}$ &
  $ 0.51\pm0.07$ &$ 0.47\pm0.08$ &$ 0.48\pm0.09$ &$ 0.55\pm0.09$ &$ 0.22\pm0.19$ \\ $M_{\scriptscriptstyle21}^{\scriptscriptstyle-}=\Omega_{z}$ &
  $ 0.06\pm0.05$ &$-0.03\pm0.05$ &$-0.06\pm0.07$ &$-0.16\pm0.06$ &$-0.53\pm0.12$ \\\hline
\end{tabular}
\end{center}
{\small\baselineskip=1.0ex\protect
The velocities $(U,V,W)_\odot$ and $\sigma_0$ are given in km s$^{-1}$, $M_{p,q}^-, p,q=1,2,3$ in km s$^{-1}$.
}
\end{table*}}

 \subsection{Rotation of the Galaxy} \label{rotat-parameters}
Linear velocity $V_0$ is the most important local parameter. In our case, the most interesting is its value $|V_0|=236.3\pm3.3$~km s$^{-1}$ (for the accepted value $R_0=8.1\pm0.1$~kpc), indicated in the last column of the table.~\ref{t1}. This estimate is unbiased, since it was found from a sample of Cepheids that satisfies the completeness condition.

\cite{Mroz19}, when analyzing about 770 Cepheids from the list of~\cite{Skowron19a} with known line-of-sight velocities, obtained the estimate $|V_0|=233.6\pm2.8$~km s$^{-1}$ (for the accepted value $R_0=8.122\pm0.031$~kpc).

\cite{Bob2021} analyzed about 800 Cepheids from the list of~\cite{Skowron19a} with known line-of-sight velocities using a nonlinear rotation model of the form~(\ref{EQ-11})--(\ref{EQ-33}) and obtained $|V_0|=240\pm3$~km s$^{-1}$ for the found value $R_0=8.27\pm0.10$~kpc.

It is also interesting to note the work of \cite{Ablimit20}, where about 3500 classical Cepheids with proper motions from the Gaia~DR2 catalog were used to construct the rotation curve of the Galaxy. The rotation curve of the Galaxy was plotted in the range of distances $R:4-19$~kpc. The circular velocity of rotation of the near-solar neighborhood was found with a very high accuracy, its value was $|V_0|=232.5\pm0.9$~km s$^{-1}$ for the accepted value $R_0=8.122\pm0.031$~kpc.

On the other hand, to form the residual velocities and analyze the kinematics of the most distant Cepheids in the sample, it is necessary to use the rotation curve~(\ref{solution-24}) or the parameters from the first columns of Table~\ref{t1}. As noted above, the rotation curve~(\ref{solution-24}) is good because it is close to flat (Fig.~\ref{f1-rotcurve}) over a very large range of distances $R$. It is applicable over the entire $R$ interval where the Cepheids of our sample are present, so there is no need to use alternative approaches. By alternative we mean, for example, a flat rotation curve ($V_0=const$), or obtained in some potential (curve~(18) in \cite{Bob2021}).

To determine the parameters of the Galaxy rotation, in addition to Cepheids, a wide variety of galactic objects are used. For example,~\cite{Dias2005} used kinematic data on 212 open star clusters primarily to estimate the parameters of the spiral pattern. In this case, the estimate $|V_0|=190$~km s$^{-1}$ was obtained for the accepted $R_0=7.5$~kpc.

In the work of~\cite{Bovy2012}, to determine the parameters of the Galaxy rotation in the distance range $R:4-14$~kpc, the line-of-sight  velocities of 3365 stars from the SDSS-III/APOGEE spectroscopic survey (Apache Point Observatory Galactic Evolution Experiment,~\cite{Wilson2010}) were used. They obtained an estimate $|V_0|=218\pm6$~km s$^{-1}$ for the found $R_0=8.1^{+1.2}_{-0.1}$~kpc.

In the work of~\cite{Lopez-Corredoira2014}, proper motions of a huge selection of red clump giants from the PPMXL catalog~\citep{Roeser2010} were used. For these stars photometric distances were determined. As a result, an estimate $|V_0|=238$~km s$^{-1}$ was obtained for the accepted $R_0=8$~kpc and a rotation curve of the Galaxy was constructed in the range of distances $R:4-16$~kpc. The rotation curve turned out to be close to the flat one over the entire analyzed $R$ range. In this case, a strong decrease in the circular velocity of galactic rotation with height was found for $|z|\approx 2$~kpc and $R\approx16$~kpc.

\cite{Galazutdinov15} concluded that the rotation curve of the thin gaseous disk of our Galaxy is Keplerian rather than flat. For this, data on distances and line-of-sight velocities obtained from observations of the interstellar lines Ca\,II H and K were used. Such observations, after calibration of~\cite{Megier09} by the trigonometric parallaxes of the Hipparcos catalog, determine a unique scale of distances to stars. According to estimates by~\cite{Megier09}, the accuracy of determining distances to OB stars by this method is about 15\%. With all this, \cite{Galazutdinov15}'s conclusion about the Keplerian nature of the rotation curve of the Galaxy contradicts the rotation curve of the Galaxy obtained in this work (which is close to the flat one).

It is interesting to note the work of~\cite{Chrobakova20}, where the rotation curve of the Galaxy was constructed using the kinematic data on stars from the Gaia\,DR2 catalog using the Jeans equation. These authors showed that in the outer region of the Galaxy the rotation curve they found in the range of distances $R:8-18$~kpc is very close to the flat one and depends little on $R$ and $z.$

\cite{Rastorguev2017}, based on data on 130 galactic masers with measured trigonometric parallaxes, found the solar velocity components
$(U_\odot,V_\odot)=(11.40,17.23)\pm(1.33,1.09)$~km s$^{-1}$, and the following values of the parameters of the Galaxy rotation curve:
  $\Omega_0=-28.93\pm0.53$~km s$^{-1}$ kpc$^{-1}$,
  $\Omega^{'}_0=3.96\pm0.07$~km s$^{-1}$ kpc$^{-2}$ and
  $\Omega^{''}_0=-0.87\pm0.03$~km s$^{-1}$ kpc$^{-3}$,  $|V_0|=243\pm10$~km s$^{-1}$ for the found value $R_0=8.40\pm0.12$~kpc.

\cite{Reid2019} found the following values for the two most important kinematic parameters for a sample of 147 masers: $R_0=8.15\pm0.15$~kpc and $\Omega_\odot=-30.32\pm0.27$~km s$^{-1}$ kpc$^{-1}$, where $\Omega_\odot=\Omega_0+V_\odot/R.$ The velocity value $V_\odot=12.24$~km s$^{-1}$ was taken from \cite{Schonrich2010}. These authors used a series expansion of the linear velocity of the Galaxy rotation.

Based on a similar approach, \cite{Hirota2020} obtained the following estimates from an analysis of 99 masers observed using the VERA program:
   $R_0=7.92\pm0.16$\,(stat.) $\pm0.3$\,(system)~kpc and
   $\Omega_\odot=-30.17\pm0.27$\,(stat.) $\pm0.3$\,(system)~km s$^{-1}$ kpc$^{-1}$, where
$\Omega_\odot=\Omega_0+V_\odot/R,$ and the velocity value $V_\odot=12.24$~km s$^{-1}$ was also taken from \cite{Schonrich2010}.

From the analysis of 788 Cepheids from the list of \cite{Mroz19} with proper motions and line-of-sight velocities from the Gaia DR2 catalog, \cite{Bob2021} found
$(U_\odot,V_\odot,W_\odot)=(10.1,13.6,7.0)\pm(0.5,0.6,0.4)$~km s$^{-1}$, and also:
 \begin{equation}
 \label{solution-2021}
 \begin{array}{cll}
       \Omega_0=-29.05\pm0.15~\hbox{km s$^{-1}$ kpc$^{-1}$},\\
   \Omega^{'}_0=3.789\pm0.045~\hbox{km s$^{-1}$ kpc$^{-2}$},\\
  \Omega^{''}_0=-0.722\pm0.027~\hbox{km s$^{-1}$ kpc$^{-3}$},\\
 \Omega^{'''}_0= 0.087\pm0.007~\hbox{km s$^{-1}$ kpc$^{-4}$},\\
            R_0=8.27\pm0.10~\hbox{kpc},
 \end{array}
 \end{equation}
where $|V_0|=240.2\pm3.2$~km s$^{-1}$ and $R_0$ was also considered as the unknown variable. In Fig.~5 from the work of \cite{Bob2021}, one can see that the rotation curve~(\ref{solution-2021}) for the formation of residual stellar velocities is suitable for distances $R$ less than 18--19~kpc.

 \subsection{Linear model parameters}
 \subsubsection{Plane $XY$}
Consider the displacement tensor describing the residual rotation around the $z$ axis. Let's denote this tensor as $M_{xy},$ since its elements are the partial derivatives of the velocities $U,V$ with respect to $x$ and $y$:
 \begin{equation}
 M_{xy}=\begin{pmatrix}
 {\partial U}/{\partial x}& {\partial U}/{\partial y}\\
 {\partial V}/{\partial x}& {\partial V}/{\partial y}\end{pmatrix}.
 \end{equation}
The elements of this tensor can be written in terms of the well-known Oort constants $(A,B,C,K)_{xy},$ which in our case describe the residual effects:
 \begin{equation}
 M_{xy}=\begin{pmatrix}
 K+C & A-B \\
 A+B & K-C \end{pmatrix}.
  \end{equation}
According to the data from the second column of Table~\ref{t1}, we have (matrix elements are given in km s$^{-1}$ kpc$^{-1}$):
 \begin{equation}
 M_{xy}=\begin{pmatrix}
 0.56_{(0.14)} & 0.16_{(0.09)} \\
 0.32_{(0.12)} & 0.22_{(0.13)}\end{pmatrix},\label{Mxy}
  \end{equation}
on the basis of which we find
$A_{xy}=0.24\pm0.08$~km s$^{-1}$ kpc$^{-1}$,  $B_{xy}=0.08\pm0.08$~km s$^{-1}$ kpc$^{-1}$,
$C_{xy}=0.17\pm0.10$~km s$^{-1}$ kpc$^{-1}$ and $K_{xy}=0.39\pm0.10$~km s$^{-1}$ kpc$^{-1}$.

In a similar approach, using for these Cepheids the proper motions from the Gaia DR2 catalog, \cite{BB2021b} found
$A_{xy}=-0.13\pm0.11$~km s$^{-1}$ kpc$^{-1}$, $B_{xy}=-0.13\pm0.11$~km s$^{-1}$ kpc$^{-1}$,
$C_{xy}=0.43\pm0.11$~km s$^{-1}$ kpc$^{-1}$ and $K_{xy}=0.25\pm0.11$~km s$^{-1}$ kpc$^{-1}$. We see that there is no particular repeatability of the results. Most likely, there are random fluctuations of the velocity field of distant Cepheids. However, $M_{11}=\partial U/\partial x$ is always significantly different from zero, as you can see from Table~\ref{t2}.

 \subsubsection{Plane $YZ$} \label{warpYZ}
Consider the displacement tensor $M_{yz}$:
 \begin{equation}
 M_{yz}=\begin{pmatrix}
 {\partial V}/{\partial y}& {\partial V}/{\partial z}\\
 {\partial W}/{\partial y}& {\partial W}/{\partial z}\end{pmatrix}.
 \end{equation}
According to the data from the first column of Table~\ref{t1} we have
 \begin{equation}
 M_{yz}=\begin{pmatrix}
 0.17_{(0.13)} & 4.72_{(1.44)} \\
 0.06_{(0.08)} & 1.55_{(1.02)}\end{pmatrix}.\label{Myz1}
  \end{equation}
If we strictly follow the rule~(\ref{B-xyz}), then here we get a negative rotation around the $x$ axis,
$M_{\scriptscriptstyle32}^{\scriptscriptstyle-}=-2.33\pm0.72$~km s$^{-1}$ kpc$^{-1}$.

In the work of \cite{BB2021b}, for a similar case, it was found
$M_{\scriptscriptstyle32}^{\scriptscriptstyle-}=-3.55\pm0.76$~km s$^{-1}$ kpc$^{-1}$. It was noted that in reality, a large positive value of the gradient $\partial V/\partial z$ is hard to believe, as it means that the galactic rotation velocity should increase with increasing $z$. In reality, exactly the opposite is the case.

On the other hand, the gradient $\partial V/\partial z~(M_{23})$ differs significantly from zero only in the first column of Table~\ref{t1}. For example, according to the data from the second column of Table~\ref{t1} we have
 \begin{equation}
 M_{yz}=\begin{pmatrix}
 ~~0.22_{(0.13)} & 3.04_{(1.67)} \\
  -0.02_{(0.08)} & 1.27_{(1.20)}\end{pmatrix},\label{Myz2}
  \end{equation}
where none of the matrix elements differs significantly from zero. Following the rule~(\ref{B-xyz}), we obtain the estimate $M_{\scriptscriptstyle32}^{\scriptscriptstyle-}=-1.53\pm0.84$~km s$^{-1}$ kpc$^{-1}$.

We can conclude that in the velocities of distant Cepheids, the residual kinematic effects in the $YZ$ plane are negligible.

 \subsubsection{Plane $XZ$} \label{warpXZ}
Here the displacement tensor $M_{xz}$ looks like this:
 \begin{equation}
 M_{xz}=\begin{pmatrix}
 {\partial U}/{\partial x}& {\partial U}/{\partial z}\\
 {\partial W}/{\partial x}& {\partial W}/{\partial z}
 \end{pmatrix}\end{equation}
First, we note the results from the last column of Table~\ref{t2}, obtained for $d\leq2$~kpc. The large value of $M_{13}=-36.30\pm8.42$~km s$^{-1}$ kpc$^{-1}$ is striking.
For this case, we have
 \begin{equation}
 M_{xz}=\begin{pmatrix}
 -0.33_{(1.01)} & -36.30_{(8.42)} \\
 -0.35_{(0.86)} & ~-4.75_{(8.30)}
  \end{pmatrix}.\label{Mxz-r2}
  \end{equation}
According to the rule~(\ref{B-xyz}), we get a large positive rotation around the $y$ axis,
$M_{\scriptscriptstyle31}^{\scriptscriptstyle-}=17.98\pm4.23$~km s$^{-1}$ kpc$^{-1}$. The value of such rotation strongly depends on the value of $M_{13}$.  As can be seen from Table~\ref{t2}, $M_{13}$ tends to zero with increasing sample radius.

For the data from the first column of table.~\ref{t2}, we have
 \begin{equation}
 M_{xz}=\begin{pmatrix}
  0.40_{(0.14)} & -1.52_{(1.20)} \\
 -0.49_{(0.08)} &  1.55_{(1.02)}
 \end{pmatrix},\label{Mxz-rall}
 \end{equation}
then we find analogs of the Oort constants for this plane
$A_{xz}= 0.24\pm0.08$~km s$^{-1}$ kpc$^{-1}$,
$B_{xz}=M_{13}^-=\Omega_y=0.52\pm0.60$~km s$^{-1}$ kpc$^{-1}$,
$C_{xz}=-1.00\pm0.60$~km s$^{-1}$ kpc$^{-1}$ and
$K_{xz}=M_{13}^+=0.39\pm0.10$~km s$^{-1}$ kpc$^{-1}$.
We see that with this approach, there is no significantly different rotation of $\Omega_y$.

As a result, we can conclude the following. Compared to the analysis of a similar sample of Cepheids with proper motions from the Gaia DR2 catalog carried out by \cite{BB2021b}, the values of the gradients  $M_{13}=\partial U/\partial z$ and $M_{23}=\partial V/\partial z$ have significantly decreased. Their values are highly depend on the residual velocities of the stars. In this work, we found that when the most distant Cepheids are included in the consideration, the values of these gradients do not differ significantly from zero.

The most interesting for analyzing the influence of the warp on the kinematics of distant Cepheids is the gradient $M_{31}=\partial W/\partial x$. It has a stable value close to $-0.5$~km s$^{-1}$ kpc$^{-1}$ under any restrictions on the sample size of Cepheids (except for the nearest ones). The $W$ velocity is independent of the galactic rotation. The most reliable estimate of the rotation around the galactic axis $y$ can be considered the value
$M_{\scriptscriptstyle31}^{\scriptscriptstyle-}=0.51\pm0.07$~km s$^{-1}$ kpc$^{-1}$, found from distant Cepheids based on a simple model of rigid body rotation (Table~\ref{t3}).

 \section{CONCLUSION}
The three-dimensional motions of a large sample of classical Cepheids of the Milky Way are analyzed. For this, the Cepheids from~\cite{Skowron19a} were used. After identifying these stars with the Gaia\,EDR3 catalog, a sample of about 2000 stars with proper motions, heliocentric distances and estimates of their ages was obtained. For some of these stars, the line-of-sight velocities are known from the Gaia\,DR2 catalog.

The parameters of the rotation curve of the Galaxy are determined on the basis of a nonlinear rotation model. In particular, it is shown that the galactic rotation curve corresponding to the solution~(\ref{solution-24}) is flat over a very wide range of distances, up to $R\sim24$~kpc. It can be recommended, for example, for obtaining the residual velocities of stars when performing spectral analysis of the residual velocities of stars and other similar works. Thus, we used the parameters~(\ref{solution-24}) to form the residual velocities of Cepheids located in a wide range of distances $R:3-18$~kpc.

The parameters of rotation of the Galaxy obtained from more near Cepheids are of interest. Thus, using stars from the region with a radius of $d=5$~kpc (this sample of Cepheids  satisfies the completeness condition), the circular linear velocity of rotation of the circumsolar neighborhood around the center of the Galaxy was found as
$V_0=236\pm3$~km s$^{-1}$ for the accepted distance $R_0=8.1\pm0.1$~kpc.

Application of the linear Ogorodnikov-Milne model to the analysis of residual velocities of Cepheids showed the presence of the gradients
$\partial U/\partial x,$ $\partial U/\partial z,$ $\partial V/\partial x,$
$\partial V/\partial z$  and $\partial W/\partial x$ significantly different from zero.  These gradients behave differently depending on the selection radius $d$ and constraints on the $R$ radius.

The most interesting is the gradient $\partial W/\partial x,$ since the velocities of $W$ are free of galactic rotation. Its value is $\sim-0.5\pm0.1$~km s$^{-1}$ kpc$^{-1}$, which can be interpreted as the positive rotation of this star system around the galactic axis $y$, i.e. rotation with angular velocity $\Omega_y$. A more accurate estimate of $\Omega_y=0.51\pm0.07$~km s$^{-1}$ kpc$^{-1}$ is obtained on the basis of a simpler model of pure rotation. In our opinion, this rotation around the $y$ axis reflect the indirect influence of some process that perturbs the vertical velocities of Cepheids.

Note that the $\partial W/\partial x$ gradient can be explained in many ways, not necessarily by deformation precession. For example, such as bending + breathing modes, fusion and interaction with satellites, changes in the amplitude of deformation (not only its precession), disequilibrium due to the violent origin of the Galactic disk, and others \citep{Wang2018a,Laporte2018}.

It is shown that on the scales under consideration ($d>12$~kpc), there is no significantly different from zero rotation around the galactic axis $x$. It was not found the volumetric expansion/contraction of the Cepheid system significantly different from zero, i.e. $K_{xyz}=0$.



 \end{document}